\documentclass[amssymb,nofootinbib,nobibnotes,aps,prd,preprintnumbers]{revtex4}

 \usepackage{here}

\usepackage[dvipdfmx]{graphicx}
\usepackage{amssymb,amsfonts,amsmath,bm,color,multirow,cases,empheq,hyperref}
\usepackage{mathrsfs}

\usepackage{enumerate}
\usepackage{ulem}
\usepackage{afterpage}

\hypersetup{%
 setpagesize=false,
 bookmarksnumbered=true,%
 bookmarksopen=true,%
 colorlinks=true,%
 linkcolor=blue,%
 citecolor=red}

\begin{document}




\title{
Asymmetric dyonic multi-centered rotating black holes
}

\author{Shinya Tomizawa}
\email{tomizawa@toyota-ti.ac.jp}

\author{Jun-ichi Sakamoto}
\email{jsakamoto@toyota-ti.ac.jp}

\affiliation{Mathematical Physics Laboratory, Toyota Technological Institute,
Hisakata 2-12-1, Nagoya 468-8511, Japan}
\date{\today}

\author{ Ryotaku Suzuki}
\email{suzuki.ryotaku@nihon-u.ac.jp}
\affiliation{Laboratory of Physics, College of Science and Technology, Nihon University,\\
Narashinodai 7-24-1, Funabashi, Chiba 274-8501, Japan}

\preprint{TTI-MATHPHYS-35}




\begin{abstract} 
We construct an exact solution in four-dimensional Einstein-Maxwell-dilaton theory, describing multi-centered rotating black holes carrying both electric and magnetic charges, obtained via dimensional reduction from five-dimensional Einstein gravity.
This  generalizes the Majumdar-Papapetrou solution to the rotating case, and extends the recent multi-centered rotating black hole solutions of Teo and Wan to configurations with unequal electric and magnetic charges. The resulting spacetimes are free of curvature singularities, conical defects, Dirac-Misner strings, and closed timelike curves, both on and outside the horizons, provided that the black holes have either aligned or anti-aligned spin orientations.
\end{abstract}

\date{\today}
\maketitle



\section{Introduction}

Exact solutions describing multi-black hole systems are of considerable interest in both astrophysics and theoretical physics, as they provide valuable insights into the behavior of black hole binaries—systems that play a central role in gravitational wave astronomy. However, obtaining such solutions remains a formidable challenge due to the lack of symmetries and the inherently dynamical nature of the problem.
Despite these difficulties, certain classes of static and stationary solutions describing black hole binaries have been identified. One of the earliest examples was presented by Israel and Khan~\cite{Israel1964}, who found an exact solution representing static, axisymmetric configurations of multiple Schwarzschild black holes aligned along a common axis of rotation. However, the gravitational attraction between the black holes in these configurations necessitates the existence of conical singularities (or struts) between them to maintain equilibrium.
As a rotational generalization of the double Schwarzschild solution, Kramer and 
Neugebauer~\cite{Kramer1980} constructed the double Kerr solution, describing the interaction of two rotating black holes. The solution reflects the delicate balance between gravitational attraction and the spin-spin repulsion arising from their angular momenta. However, despite this interplay, conical singularities persist in the intermediate region, implying that exact vacuum solutions of static or stationary multi-black hole systems necessarily contain singularities.

\medskip
On the other hand, the inclusion of electric charge changes this situation. A well-known example is the Majumdar-Papapetrou solution~\cite{Majumdar:1947eu,Papapetrou}, which provides an exact static multi-black hole solution to the Einstein-Maxwell equations. In this configuration, the mutual gravitational attraction between the black holes is exactly balanced by the electrostatic repulsion due to their charges, allowing for static equilibrium without the need for conical singularities. Israel and Wilson~\cite{Israel:1972vx}, along with Perj\'es~\cite{Perjes:1971gv}, extended the Majumdar-Papapetrou solution to include rotation, obtaining a class of stationary solutions within the Einstein-Maxwell theory. 
It was later shown, however, that these solutions represent configurations of naked singularities rather than genuine black holes. This result suggests that equilibrium configurations of rotating charged black holes cannot be realized within the framework of Einstein-Maxwell theory alone.
More recently, Teo and Wan~\cite{Teo:2023wfd} succeeded in constructing a new class of exact regular solutions describing multi-centered spinning black holes in 5D Kaluza-Klein theory. Upon dimensional reduction, these solutions correspond to balanced configurations of arbitrarily many dyonic rotating black holes in the 4D Einstein-Maxwell-dilaton theory. Each black hole in the solution is characterized by its own mass, angular momentum, equal electric and magnetic charges, and position. Furthermore, in the limit where all spin angular momenta vanish, the solution reduces to the Majumdar-Papapetrou solution, as the scalar (dilaton) field also vanishes in this limit.

\medskip
 The exact solutions for such Kaluza-Klein black holes were obtained by many researchers.
Specifically, in 5D Kaluza-Klein theory, a black hole solution carrying a Kaluza-Klein (K-K) electric charge but no magnetic charge-commonly referred to as the boosted Schwarzschild string-was first investigated by Chodos and Detweiler~\cite{Chodos-Detweiler}. 
Shortly thereafter, Frolov, Zel'nikov, and Bleyler~\cite{FZB} extended this construction to include rotation.
The first genuinely non-trivial configuration, in which a Kerr black hole is twisted along the compact Kaluza-Klein circle, was subsequently obtained by Dobiasch and Maison~\cite{DM} through a transformation in the framework of the non-linear sigma model. 
Their solution was later analyzed in detail by Gibbons and Wiltshire~\cite{GW}, Pollard~\cite{Pollard}, and Gibbons and Maeda~\cite{Gibbons-Maeda}.
A more general class of rotating Kaluza-Klein black hole solutions, incorporating both K-K electric and magnetic charges (interpreted as dyonic rotating black holes from the 4D viewpoint), was constructed by Rasheed~\cite{Rasheed}, who employed an $SL(3,\mathbb{R})$ transformation acting on the Kerr string background characterized by a mass parameter  and a rotation parameter. Independently, the same family of solutions was also obtained by Larsen~\cite{Larsen:1999pp,Larsen:1999pu}.

\medskip
As shown by Maison~\cite{Maison:1979kx}, the 5D vacuum Einstein theory with two commuting Killing vectors can be reduced to a 3D gravity-coupled nonlinear sigma model with an $SL(3,\mathbb{R})$ target space.
In general, solving the resulting equations is challenging, as standard solitonic techniques such as the inverse scattering method cannot be applied due to the lack of a third Killing isometry.
However, Cl\'ement~\cite{Clement:1985gm,Clement:1986bt}  showed that a special class of solutions can be constructed from flat 3D Euclidean space and two harmonic functions.
The slowly rotating extremal limit of the Rasheed-Larsen solution falls  within this class.
By generalizing the two harmonic functions to multi-centered configurations, Teo and Wan~\cite{Teo:2023wfd} constructed a multi-rotating Kaluza-Klein black hole solution that, after dimensional reduction to 4D, describes an asymptotically flat rotating black hole with equal electric and magnetic charges. 
The purpose of this paper is to extend the Teo-Wan solution to the more general case with unequal electric and magnetic charges in 5D Kaluza-Klein theory.
The resulting spacetime is free of curvature singularities, conical defects, Dirac-Misner strings, and closed timelike curves, on and outside the horizons, provided that the black holes have either aligned or anti-aligned spin orientations.

\medskip
We briefly outline the organization of this paper.
In Section~\ref{sec:TW}, we review the multi-centered rotating black hole solutions of Teo and Wan in 5D Kaluza-Klein theory.
In Section~\ref{sec:formalism}, we present the non-linear sigma model framework developed by Maison and show how the 5D Einstein equations with two commuting Killing vectors reduce to a 3D gravity-coupled sigma model. We also provide the necessary equations for our analysis, following the work of Cl\'ement.
In Section~\ref{sec:solution}, we see that the slowly rotating extremal limit of the Rasheed-Larsen solution falls within the class of solutions constructed from two harmonic functions with a single point source. 
By extending these harmonic functions from a single point source to multiple point sources, we construct a more general class of multi-centered rotating black hole solutions, thereby extending the Teo-Wan solutions to the case of unequal electric and magnetic charges
In Section~\ref{sec:anaysis}, we analyze the properties of our solutions in detail, including the structure of the horizons, their asymptotic behavior, regularity conditions, and the absence of closed timelike curves.
We also discuss the parameter region corresponding to physical solutions and examine various physical limits in which our solutions reduce to previously known configurations.
Finally, Section~\ref{sec:summary} is devoted to a summary and discussion.

\section{The Teo-Wan solutions}\label{sec:TW}

\medskip

Let us start with the 5D Kaluza-Klein theory, in which the metric can be written as
\begin{eqnarray}
ds^2=e^{-\frac{2\phi}{\sqrt{3}}}(dx^5+{\bm A})^2+e^{\frac{\phi}{\sqrt{3}}}g_{\mu\nu} dx^\mu dx^\nu,
\end{eqnarray}
where the function $\phi$, the component $A_\mu$ of the $1$-form ${\bm A}=A_\mu dx^\mu $ and the 4D metric $g_{\mu\nu}$  ($\mu,\nu=0,\ldots,3$)  do not depend on the fifth spatial coordinate $x^5$ which has a period of $2\pi R_{KK}$. 
As is well-known, the dimensional reduction of the 5D Einstein theory leads the 4D  Einstein-Maxwell-dilaton theory, described by the action 
\begin{eqnarray}
S=\int d^4x\sqrt{-g} \left( R-\frac{1}{2}\partial_\mu \phi\partial^\mu \phi -\frac{1}{4}e^{-\sqrt{3} \phi} F^2 \right),
\end{eqnarray}
where $R$ is the Ricci scalar of the 4D metric, $g={\rm det}(g_{\mu\nu})$, and  $F_{\mu\nu}:=\partial_\mu A_\nu-\partial_\nu A_\mu$ is the field strength.
From this, the fields equations, the Einstein equation for the 4D metric $g_{\mu\nu}$, the equations for the gauge potential $A_\mu$ and the scalar fields $\phi$ can be written as, respectively, 
\begin{eqnarray}
R_{\mu\nu}=\frac{1}{2}\partial_\mu \phi \partial_\nu \phi +\frac{1}{2}e^{-\sqrt{3} \phi} \left(F_{\mu\rho}F_\nu{}^\rho-\frac{1}{4} g_{\mu\nu} F^2 \right),
\end{eqnarray}
\begin{eqnarray}
\nabla_\mu\left(e^{-\sqrt{3} \phi}F^{\mu\nu}\right)=0,
\end{eqnarray}
\begin{eqnarray}
\nabla_\mu\nabla^\mu\phi=-\frac{\sqrt{3}}{4} e^{-\sqrt{3} \phi} F^2.
\end{eqnarray}

The 4D metric, the gauge potential and the scalar field of the multi-centered rotating black hole solution, constructed by Teo and Wan~\cite{Teo:2023wfd}, can be written, respectively,  as
\begin{eqnarray}
ds^2_{(4)}&=&g_{\mu\nu}dx^\mu dx^\nu=-(H_+H_-)^{-\frac{1}{2}}(dt+{\bm \omega^0})^2+(H_+H_-)^{\frac{1}{2}}d{\bm x}\cdot d{\bm x},\\
{\bm A}&=&\frac{\sqrt{2}}{H_-}\left[-[(1+f)f-2g]dt+(1+f){\bm\omega^0}+ H_- \tilde{\bm \omega}^5\right],\\
\phi&=&\frac{\sqrt{3}}{2}\ln \frac{H_+}{H_-},
\end{eqnarray}
where $d{\bm x}\cdot d{\bm x}$ is the metric of the 3D Euclid space ${\mathbb E}^3$ with ${\bm x}=(x,y,z)$. 
The functions $H_\pm$, one-forms ${\bm \omega}^0$, $\tilde{\bm \omega}^5$ on ${\mathbb E}_3$ are given by
\begin{eqnarray}
H_\pm&=&(1+f)^2\pm 2g,\\
{\bm \omega}^0&=&-\sum_{i=1}^N\frac{2J_i[(y-y_i)dx-(x-x_i)dy]}{|{\bm x}-{\bm x}_i|^3},\\
\tilde{\bm \omega}^5&=&\sqrt{2}\sum_{i=1}^N\frac{M_i(z-z_i)[(y-y_i)dx-(x-x_i)dy]}{|{\bm x}-{\bm x}_i|[(x-x_i)^2+(y-y_i)^2]},
\end{eqnarray}
with the two harmonic functions, $f$ and $g$, having point sources at the positions ${\bm x}={\bm x_i}:=(x_i,y_i,z_i)$ ($i=1,\ldots,N$) on ${\mathbb E}^3$,
\begin{eqnarray}
f=\sum_{i=1}^N\frac{M_i}{|{\bm x}-{\bm x_i}|},\quad 
g=\sum_{i=1}^N\frac{J_i(z-z_i)}{|{\bm x}-{\bm x_i}|^3}.
\end{eqnarray}
This solution describes an asymptotically flat, stationary multi-centered rotating dyonic black holes, with each having an extremal horizon.
The $i$-th black hole at the position ${\bm x}={\bm x}_i$ on ${\mathbb E}^3$ carries the mass $M_i$, spin angular momentum $J_i$, equal electric and magnetic charges  $Q_i$ and $P_i$, respectively, given by $Q_i=P_i=M_i/\sqrt{2}$.  
Furthermore, the regularity of the metric on the horizon requires the  condition
\begin{eqnarray}
|J_i|<\frac{M_i^2}{2}, \label{eq:regularity}
\end{eqnarray}
since the horizon area, $4\pi \sqrt{M_i^4-4J_i^2}$, vanishes if this is saturated. 
Moreover, as shown in Ref.~\cite{Teo:2023wfd}, under the condition~(\ref{eq:regularity}), the spacetime is free from closed timelike curves on and outside the horizon. 
At the limit as $J_i\to 0$ for all $i$, the scalar field $\phi$ vanishes, and consequently, the Majumdar-Papapetrou solution describing static multiple dyonic black holes is restored.

\section{5D Kaluza-Klein theory and non-linear sigma model}\label{sec:formalism}
Since we wish to consider stationary solutions, in addition to the Killing vector $\partial/\partial x^5$, we assume the existence of a timelike Killing vector field $\partial/\partial t$. With these two commuting Killing vectors, the 5D Einstein equations reduce to a 3D system of gravity coupled to scalar fields~\cite{Maison:1979kx}. We then review how this system of scalar fields can be described by a nonlinear sigma model. In general, solving the resulting equations is challenging because the scalar fields are coupled to 3D gravity. However, under the assumption that the 3D geometry is flat, Cl\'ement~\cite{Clement:1985gm,Clement:1986bt} showed that a special class of solutions can be constructed using two harmonic functions. We also present the necessary equations for our analysis, following Cl\'ement's formulation.

\subsection{5D Einstein equation with two commuting Killing vectors}

Let $\xi_a \ (a=0,5)$ be two mutually commuting Killing vector fields,  so that $[\xi_a,\xi_b]=0$, ${\cal L}_{\xi_a} g=0$.  Then, introducing the coordinates $x^a$ as Killing parameters of $\xi_a$ 
(so that $\xi_a = \partial/\partial x^a$), one can express the metric $g$ as     
\begin{eqnarray}
ds^2 = \lambda_{ab}(dx^a+\omega^a{_i}dx^i)(dx^b+\omega^b{_j}dx^j) 
      +|\tau|^{-1}h_{ij}dx^idx^j \,, \label{eq:5D metric}
\end{eqnarray}
where the functions $\tau:=-{\rm det}(\lambda_{ab})$, $\omega^a{_i}$, $h_{ij}$ ($i=1,2,3$) are independent of 
the coordinates~$x^a$.  
We also introduce the twist one-forms by 
\begin{equation} 
v_a=*(\xi_0\wedge \xi_5\wedge d\xi_a) \,. 
\end{equation}
Then, we can write the exterior derivative of $V_a$ as 
\begin{eqnarray}
dv_a &=& 2*(\xi_0\wedge \xi_5\wedge R(\xi_a)),
 \end{eqnarray}
where $R(\xi_a)$  is the Ricci one-form. 
Therefore, since $R(\xi_a)=0$ from the vacuum Einstein equation, there exists locally the twist potentials $V_a$ that satisfy 
\begin{eqnarray}
 dV_a=v_a
\label{eq:twistpotential} 
\end{eqnarray} 
which can be written as

\begin{eqnarray}
  \partial_kV_{a}
 =\tau \sqrt{|h|} \lambda_{ab} \varepsilon_{kij}  h^{im}h^{jn}\partial_m \omega^b{}_n.\label{eq:domega}
\end{eqnarray}
Then, the vacuum Einstein equations can be expressed as
the field equations for the five scalar fields, $\{\lambda_{ab},V_a\}$, 
\begin{eqnarray}
\Delta_h \lambda_{ab}&=&\lambda^{cd} h^{ij}\frac{\partial \lambda_{ac}}{\partial x^i} \frac{\partial \lambda_{bd}}{\partial x^j}+\tau^{-1} h^{ij}\frac{\partial V_a}{\partial x^i} \frac{\partial V_b}{\partial x^j},\label{eq:eom1}\\
\Delta_h V_{a}&=& \tau^{-1} h^{ij}\frac{\partial \tau}{\partial x^i} \frac{\partial V_a}{\partial x^j}+\lambda^{bc} h^{ij}\frac{\partial \lambda_{ab}}{\partial x^i} \frac{\partial V_c}{\partial x^j},\label{eq:eom2}
\end{eqnarray}
and the Einstein equations for the 3D metric $h_{ij}$, which is coupled with the five scalar fields,
\begin{eqnarray}
R^h_{ij} &=&  \frac{1}{4} \lambda^{ab}\lambda^{cd}
              \frac{\partial \lambda_{ac}}{\partial x^i }  \frac{\partial \lambda_{bd}}{\partial x^j } 
   + \frac{1}{4}\tau^{-2}\frac{\partial \tau}{\partial x^i} \frac{\partial \tau}{\partial x^j }   
    -\frac{1}{2}\tau^{-1}\lambda^{ab} \frac{\partial V_a}{\partial x^i }\frac{\partial V_b}{\partial x^j },       
\label{eq:Rij}
\end{eqnarray} 
where $\Delta_h$ is the Laplacian and $R^h_{ij}$ denotes the Ricci tensor with respect to $h_{ij}$, respectively.

\medskip
\subsection{$SL(3,{\mathbb R})$ nonlinear sigma model}
Thus, as a consequence of the existence of two isometries, we have five scalar fields $\lambda_{ab},V_a$ 
$(a=0,5)$, which we denote collectively by coordinates $\Phi^A=(\lambda_{ab},V_a)$. 
Then, we can find that the equations of motion, Eqs.~(\ref{eq:eom1}), (\ref{eq:eom2}) and (\ref{eq:Rij})
are derived from the following action for sigma-model $\Phi^A$ coupled with three-dimensional gravity with respect to the metric $h_{ij}$, 
\begin{eqnarray}
S=\int\left(R^h
 -G_{AB}\frac{\partial \Phi^A}{\partial x^i}
 \frac{\partial \Phi^B}{\partial x^j}h^{ij}\right)\sqrt{|h|}d^3x \,,\label{eq:action}
\end{eqnarray}
where the target space metric, $G_{AB}$, is given by 
\begin{eqnarray}
G_{AB}d\Phi^Ad\Phi^B 
&=& \frac{1}{4}{\rm Tr}(\lambda^{-1}d\lambda\lambda^{-1}d\lambda )
   + \frac{1}{4}\tau^{-2}d\tau^2 
    -\frac{1}{2}\tau^{-1}v^T\lambda^{-1}v,
\end{eqnarray}
with $\lambda=(\lambda_{ab})$, $v=(v_0,v_5)^T$ and $v=dV$.
It can be actually shown that varying the action by $h_{ij}$ derives the equation~(\ref{eq:Rij}),
\begin{eqnarray}
R^h_{ij} &=& G_{AB}\frac{\partial \Phi^A}{\partial x^i} 
                 \frac{\partial \Phi^B}{\partial x^j},  
      \end{eqnarray} 
and also varying the action by $\Phi^A$ can derive the equations~(\ref{eq:eom1}) and (\ref{eq:eom2}),
\begin{eqnarray}
\Delta_h\Phi^A+h^{ij}\Gamma^A_{BC}\frac{\partial \Phi^B}{\partial x^i} 
                                  \frac{\partial \Phi^C}{\partial x^j} 
  =0, \label{eq:harmonic_map}
\end{eqnarray}
where  $\Gamma^A_{BC}$ is the Christoffel symbol with respect to $G_{AB}$.

\subsection{Coset matrix}
Maison~\cite{Maison:1979kx} showed that the action~(\ref{eq:action}) for the five scalar fields $\Phi^A$ is invariant under the global $SL(3,{\mathbb R})$ transformation, introducing the $SL(3,{\mathbb R})$ matrix $\chi$ defined by the $3\times3$ matrix:
\begin{eqnarray}
\chi= \left(
  \begin{array}{ccc}
  \displaystyle \lambda_{ab}-\frac{V_ aV_b^T}{\tau} &  \displaystyle \frac{V_a}{\tau}\\
   \displaystyle  \frac{V_b^T}{\tau}                     & \displaystyle  -\frac{1}{\tau}
    \end{array}
 \right) \,, \label{eq:chidef}
\end{eqnarray}
which is symmetric, $\chi^T=\chi$, and unimodular, $\det(\chi)= 1$.
Then, the action (\ref{eq:action}) can be expressed in terms of the matrix $\chi$ as
\begin{eqnarray}
S=\int\left(R^h
 - \frac{1}{4}h^{ij} {\rm tr}(\chi^{-1}\partial_i\chi \chi^{-1}\partial_j\chi)
\right)\sqrt{|h|}d^3x \,,
\end{eqnarray}
which  is invariant under the transformation
\begin{eqnarray}
\chi\to \chi'=g \chi g^T,\quad h\to h \label{eq:sl3r}
\end{eqnarray}
with $g\in SL(3,{\mathbb R})$.
The equations of motion~(\ref{eq:Rij}), (\ref{eq:eom1}) and  (\ref{eq:eom2})  can be written, in terms of $\chi$, as
\begin{eqnarray}\label{eq:eom}
&&d\star_h (\chi^{-1} d\chi)=0, \label{eq:eomb}\\
&&R^h_{ij}=\frac{1}{4}{\rm tr}(\chi^{-1}\partial_i\chi \chi^{-1}\partial_j\chi).\label{eq:eom2b}
\end{eqnarray}
Thus, the existence of two commuting Killing vector
fields reduces the 5D vacuum Einstein theory to a 3D non-linear sigma model with a $SL(3,{\mathbb R})$ target space symmetry. 
If both two Killing vectors are spacelike, it is described by the $SL(3,{\mathbb R})/SO(3)$ sigma model coupled to gravity, while if one of the two Killing vectors is timelike, the symmetry is replaced with $SL(3,{\mathbb R})/SO(2,1)$.

\subsection{Asymptotically flat solutions of  Cl\'ement}
Since the two equations~(\ref{eq:eomb}) and (\ref{eq:eom2b}) are mutually coupled, solving them is not straightforward.
However, when the 3D metric $h_{ij}$ is a Euclid space metric  ${\mathbb E}^3$,
\begin{eqnarray}
h_{ij}dx^idx^j=d{\bm x}\cdot d{\bm x}, \label{eq:h}
\end{eqnarray}
with the position vector ${\bm x}=(x,y,z)$ on ${\mathbb E}^3$, they can be simplified to
\begin{eqnarray}
&&\partial_i (\chi^{-1} \partial^i\chi)=0,\label{eq:eomc}\\
&&{\rm tr}(\chi^{-1}\partial_i\chi \chi^{-1}\partial_j\chi)=0.\label{eq:eom2c}
\end{eqnarray}
The general solution depending on two potentials originally obtained by Cl\'ement~\cite{Clement:1986bt,Clement:1985gm}, is given by
\begin{eqnarray}
\chi=\eta e^{fA}e^{gA^2}, \label{eq:chi}
\end{eqnarray}
where $f$ and $g$ are harmonic functions on  ${\mathbb E}^3$, $\eta$ and $A$ are $3\times 3$ constant matrices. 
To ensure that the 4D metric $ds^2_{(4)}=g_{\mu\nu}dx^\mu dx^\nu$ is asymptotically flat, the matrix $\eta$ must be chosen as
\begin{eqnarray}
\eta=
\begin{pmatrix}
-1  &  0& 0 \\
  0&1 &  0   \\
0  & 0 &-1  
\end{pmatrix},
\end{eqnarray}
if  the two harmonic functions vanish at infinity. 
If the matrix $A$ satisfies
\begin{eqnarray}
A^T=\eta A \eta, \quad {\rm tr}(A)=0, \quad 
{\rm tr}(A^2)=0,
\end{eqnarray}
then the matrix $\chi$ has the required properties: it is symmetric ($\chi^T=\chi$), unimodular (${\rm det\ \chi}=1$) and it satisfies the constraint (\ref{eq:eom2c}).

\section{Multi-centered rotating black holes with unequal magnetic and electric charges}\label{sec:solution}

Rasheed and Larsen~\cite{Rasheed,Larsen:1999pp} constructed the exact solution describing the most general dyonic rotating black holes in 5D Kaluza-Klein theory. This solution possesses two distinct branches of the extremal limit: the slowly rotating extremal limit and the fast rotating extremal limit. In particular, the former falls within the class of Cl\'ement’s solutions, whereas the latter does not. The slowly rotating extremal limit can be expressed in terms of the solution with two harmonic functions, Eq.~(\ref{eq:chi}), each associated with a single point source. In this section, by generalizing these harmonic functions to multiple point sources, we construct an exact solution describing multi-centered rotating dyonic black holes with unequal electric and magnetic charges in 4D Einstein-Maxwell-dilaton theory. This extends the construction of Ref.~\cite{Teo:2023wfd}, where the charges were taken to be equal.

\subsection{The slowly rotating extremal limit of the Rasheed-Larsen solution}
The 4D metric, gauge field and scalar field  of the Rasheed-Larsen solution~\cite{Rasheed,Larsen:1999pp} is given, respectively,  by
\begin{eqnarray}
ds^2_{(4)}=-\frac{H_3}{\sqrt{H_1H_2}}(dt+{\bm B})^2+\sqrt{H_1H_2}\left(\frac{dr^2}{\Delta}+d\theta^2+\frac{\Delta}{H_3}\sin^2\theta d\phi^2\right),
\end{eqnarray}
\begin{eqnarray}
{\bm B}&=&\sqrt{pq}\frac{(pq+4m^2)r-m(p-2m)(q-2m)}{2m(p+q)H_3}a\sin^2\theta d\phi,\\
{\bm A}&=& -\left[2Q\left(r+\frac{p-2m}{2}\right)+\sqrt{\frac{q^3(p^2-4m^2)}{4m^2(p+q)}}a\cos\theta \right]H_2^{-1}dt\notag\\
            &&-\left[ 2P(H_2+a^2\sin^2\theta)\cos\theta+\sqrt{\frac{p(q^2-4m^2)}{4m^2(p+q)^3}}\right.  \notag\\
            &&\times \left\{ (p+q)(pr-m(p-2m))+q(p^2-4m^2)      \right\}a\sin^2\theta  \biggr]H_2^{-1}d\phi,\\
     e^{-2\phi}&=&\sqrt{\frac{H_2}{H_1}},
\end{eqnarray}
where
\begin{eqnarray}
H_1&=&r^2+a^2\cos^2\theta+r(p-2m)+\frac{p(p-2m)(q-2m)}{2(p+q)}-\frac{p\sqrt{(q^2-4m^2)(p^2-4m^2})}{2m(p+q)}a\cos\theta,\\
H_2&=&r^2+a^2\cos^2\theta+r(q-2m)+\frac{q(p-2m)(q-2m)}{2(p+q)}+\frac{q\sqrt{(q^2-4m^2)(p^2-4m^2})}{2m(p+q)}a\cos\theta,\\
H_3&=&r^2+a^2\cos^2\theta -2mr,\\
\Delta&=&r^2+a^2-2mr.
\end{eqnarray}
This solution is characterized by four parameters $(m,a,q,p)$, which are related to the physical mass $M$, angular momentum $J$, electric charge $Q$ and magnetic charge $P$ through
\begin{eqnarray}\label{eq:charge}
M=\frac{p+q}{4},\quad 
J=\frac{\sqrt{pq}(pq+4m^2)}{4m(p+q)}a,\quad 
Q^2=\frac{q(q^2-4m^2)}{4(p+q)},\quad
P^2=\frac{p(p^2-4m^2)}{4(p+q)}.
\end{eqnarray}
In order that the solutions have a regular horizon, they must always satisfy the following bound:
\begin{eqnarray}
\left(\frac{P}{2M}\right)^{\frac{2}{3}}+\left(\frac{Q}{2M}\right)^{\frac{2}{3}}\le 1. \label{eq:bound}
\end{eqnarray}
The solution has two distinct branches with extremal horizons:

\begin{enumerate}
\item[(1)] Slowly rotating extremal limit:  $|J|<|PQ|$:

This limit is achieved by taking the limit $m\to 0,a\to 0$ with $j:=a/m$ fixed. 
This solution is parameterized by the three parameters $p,q$ and $j$.
The solution  saturates the bound as
\begin{eqnarray}
\left(\frac{P}{2M}\right)^{\frac{2}{3}}+\left(\frac{Q}{2M}\right)^{\frac{2}{3}}= 1.
\end{eqnarray}
Although the angular momentum $J$ does not vanish,  the angular velocity of the horizon vanishes, and therefore no ergoregion exist. 
The entropy is given by
\begin{eqnarray}
S=2\pi \sqrt{P^2Q^2-J^2}.
\end{eqnarray}

\item[(2)] Fast rotating extremal limit: $|J|>|PQ|$:

This limit is obtained by taking $a \to m$.
The resulting solution is parameterized by three quantities, $p$, $q$, and $m$.
Although the explicit relations among the physical charges $M$, $J$, $P$, and $Q$ are not easy to express, they satisfy the inequality
\begin{eqnarray}
\left(\frac{P}{2M}\right)^{\frac{2}{3}}+\left(\frac{Q}{2M}\right)^{\frac{2}{3}}< 1.
\end{eqnarray}
Unlike the slowly rotating limit, the angular velocity of the horizon does not vanishes, and therefore an ergoregion exist around the horizon. 
The entropy is given by
\begin{eqnarray}
S=2\pi \sqrt{J^2-P^2Q^2}.
\end{eqnarray}

\end{enumerate}

The slowly rotating limit corresponds to the solution of Eqs.~(\ref{eq:eomc}) and (\ref{eq:eom2c})  with (\ref{eq:h}), whereas the fast rotating limit does not these solution.
Therefore, in what follows, we restrict our attention to the slowly rotating limit only.
In the slowly rotating extremal limit, the 5D metric is simplified as
\begin{eqnarray}
ds^2_{(5)}&=&\frac{H_2}{H_1}\left[dx^5-\left\{2\left(r+\frac{p}{2}\right)-pj \cos\theta \right\}\frac{Q}{H_2}dt-\left\{2H_2\cos\theta -q\left(r+\frac{pq}{p+q}\right)j\sin^2\theta \right\}\frac{P}{H_2}d\phi\right]^2\notag\\
&&-\frac{r^2}{H_2}\left(dt+ \frac{2jPQ\sin^2\theta d\phi}{r} \right)^2+H_1\left(\frac{dr^2}{r^2}+d\theta^2+\sin^2\theta d\phi^2\right),\label{eq:RLU}
 \end{eqnarray}
with the two functions
 \begin{eqnarray}
H_{1}&=&r^2+pr+\frac{p^2q(1+j\cos\theta)}{2(p+q)},\\
H_{2}&=&r^2+qr+\frac{q^2p(1-j\cos\theta)}{2(p+q)}.
\end{eqnarray}
The 4D metric, the gauge and scalar fields are written as
\begin{eqnarray}
 ds^2_{(4)}&=&-\frac{r^2}{(H_1H_2)^{1/2}}(dt+{\bm \omega^0})^2+\frac{(H_1H_2)^{1/2}}{r^2} [dr^2+r^2(d\theta^2+\sin^2\theta d\phi^2)],\\
 {\bm A}&=&-\left[2\left(r+\frac{p}{2}\right)-pj \cos\theta \right]\frac{Q}{H_2}dt-\left[2H_2\cos\theta -q\left(r+\frac{pq}{p+q}\right)j\sin^2\theta \right]\frac{P}{H_2}d\phi,\\
 e^{\frac{2\phi}{\sqrt{3}}}&=&\frac{H_2}{H_1}.
\end{eqnarray}
The physical charges~(\ref{eq:charge}) are written as
\begin{eqnarray}
M=\frac{p+q}{4},\quad J=j PQ, \quad P^2=\frac{p^3}{4(p+q)},\quad Q^2=\frac{q^3}{4(p+q)},
\end{eqnarray}
which we can confirm that saturate the bound~(\ref{eq:bound}).

\medskip
This  slowly rotating extremal solution  belongs to the class of solutions  (\ref{eq:chi}) depending on two harmonic functions of Cl\'ement, where the corresponding harmonic functions $f$, $g$ and the matrix $A$ are given by
\begin{eqnarray}
f=\frac{M}{r}, \quad g=\frac{JM^2}{2PQ}\frac{\cos\theta }{r^2}, \label{eq:fg1}
\end{eqnarray}
\begin{eqnarray}
A=4
\begin{pmatrix}
   -\frac{q}{p+q} &  (\frac{q}{p+q} )^{\frac{3}{2}} & 0 \\
  - (\frac{q}{p+q} )^{\frac{3}{2}} & -\frac{p-q}{p+q} &   (\frac{p}{p+q} )^{\frac{3}{2}}  \\
  0 &  -(\frac{p}{p+q} )^{\frac{3}{2}} &    \frac{p}{p+q}
\end{pmatrix}
. \label{eq:A}
\end{eqnarray}
Below we show that under the assumption that the 3D metric is flat, $h_{ij}dx^idx^j=d{\bm x}\cdot d{\bm x}$, the coset matrix $\chi=\eta e^{Af}e^{A^2g}$ with (\ref{eq:fg1}) and (\ref{eq:A}) reproduces the slowly rotating extremal limit of the Rasheed-Larsen solution.
Since two parameters $p$ and $q$ (or $P$ and $Q$) are not independent, it is useful to introduce a new parameter $\alpha$ defined by
\begin{eqnarray}
\cos^2\alpha=\frac{p}{p+q},\quad \sin^2\alpha=\frac{q}{p+q},
\end{eqnarray}
where it is useful to know that  $(p,q)$ or $(P,Q)$ can be expressed as 
\begin{eqnarray}
(p,q)=(4M\cos^2\alpha,4M\sin^2\alpha),\quad  (P,Q)=(2M\cos^3\alpha,2M\sin^3\alpha).
\end{eqnarray}
From  (\ref{eq:fg1}) and (\ref{eq:A}), the coset matrix $\chi$ takes the form

\begin{eqnarray}
\setlength{\arraycolsep}{1pt}  
\chi
=
{\tiny
\begin{pmatrix}
 -8 s^4 c^2 f^2+4 s^2 f-16 c^6 g+32 c^4 g-16 c^2 g-1 & 4 s^3 \left(2 c^2 f^2+4 c^2 g-f\right) & -8 c^3s^3\left(f^2+2 g\right) \\
 4s^3\left(2 c^2 f^2+4 c^2 g-f\right) & -8s^2 c^2 f^2+\left(4-8 c^2\right) f+16 c^4 g-16 c^2 g+1 & 4 c^3 \left(2 s^2f^2+4 s^2 g+f\right) \\
 -8 c^3s^3\left(f^2+2 g\right) & 4 c^3 \left(2 s^2f^2+4s^2 g+f\right) & 8 c^6 \left(f^2+2 g\right)-8 c^4 \left(f^2+2 g\right)-4 c^2 f-1 \\
\end{pmatrix}
}
\end{eqnarray}
with $(c,s):=(\cos\alpha,\sin\alpha)$.
Hence, from Eq.~(\ref{eq:chidef}), we can read off the conformal factor $\tau$ and the scalar fields $(\lambda_{ab},V_a)$ as
\begin{eqnarray}
\tau&=&\frac{1}{8 s^2c^4 \left(f^2+2 g\right)+4 c^2 f+1},\\
\lambda_{00}&=&-\tau[-8 s^2 c^2 f^2+\left(8 c^2-4\right) f+16c^2s^2g+1],\\
\lambda_{05}&=&-4\tau s^3 \left(2 c^2 f^2+f-4 c^2 g\right),\\
\lambda_{55}&=&-\tau \left[-8 s^4 c^2 f^2-4s^2 f+16c^2s^4g-1\right] ,\\
V_0&=& -8\tau c^3 s^3 \left(f^2+2 g\right), \label{eq:V0}\\
V_5&=& 4\tau( 2 s^2 f^2 + f +4 s^2g). \label{eq:V5}
\end{eqnarray}
From Eqs.~(\ref{eq:V0}) and (\ref{eq:V5}), the 1-forms ${\bm \omega}^0$ and ${\bm \omega}^5$ can be expressed as
\begin{eqnarray}
{\bm \nabla} \times {\bm \omega}^0&=&-2\sin^32\alpha {\bm \nabla} g \label{eq:omega0a},\\
{\bm \nabla} \times {\bm \omega}^5&=&-4\cos^3\alpha ({\bm \nabla} f+4\sin^2\alpha {\bm \nabla}g ).\label{eq:omega5}
\end{eqnarray}
If we define $\tilde {\bm \omega}^5:={\bm \omega^5}-(\sin\alpha)^{-1} \ {\bm \omega^0}$, then
\begin{eqnarray}
{\bm \nabla} \times \tilde {\bm \omega}^5=-4\cos^3\alpha {\bm \nabla}  f.  \label{eq:omega5t}
\end{eqnarray}
From Eq.~(\ref{eq:fg1}),  these can be solved as
\begin{eqnarray}
{\bm \omega}^0&=&\frac{2J}{r}\sin^2\theta d\phi=-\frac{2J}{r^3}[ydx-xdy],\\
 \tilde{\bm \omega}^5&=& -2P\cos\theta d\phi=\frac{2Pz}{r}\frac{ydx-xdy}{x^2+y^2}.
\end{eqnarray}
Thus, in terms of the spherical coordinates $(r,\theta,\phi)$, defined by $(x,y,z)=(r\sin\theta\cos\phi,r\sin\theta\sin\phi,r\cos\theta)$,  the 5D metric~(\ref{eq:5D metric})  can be obtained as
\begin{eqnarray}
ds^2&=&\frac{H_-}{H_+}
\left[dx^5-\frac{dt}{\sin\alpha}+\frac{1+f \sin^22\alpha}{\sin\alpha\  H_-}\left(dt+\frac{2J}{r}\sin^2\theta d\phi\right)
-2P\cos\theta d\phi \right]^2\notag\\
&&-\frac{1}{H_-}\left (dt+\frac{2J}{r}\sin^2\theta d\phi\right )^2+H_+\left[dr^2+r^2(d\theta^2+\sin^2\theta d\phi^2)\right],
\end{eqnarray}
where the functions $H_\pm$ are given  by
\begin{eqnarray}
H_\pm=1+2f+\sin^22\alpha\ f^2+2\sin^22\alpha \cos2\alpha\ g\pm (2\sin^22\alpha \ g+2\cos 2\alpha\ f + \cos 2\alpha \sin^2 2\alpha\ f^2).\label{eq:hpm}
\end{eqnarray}
This coincides with the metric~(\ref{eq:RLU}) corresponding to the slowly rotating limit of the Rasheed-Larsen solution. 
It should be emphasized, however, that in contrast to this case, the fast rotating limit does not belong to Cl\'ement's class of solutions, since the three-dimensional metric $h_{ij}$ is not flat.

\subsection{Multi-centered rotating black hole solutions}

We replace the harmonic functions $f$ and $g$ in Eq.(\ref{eq:fg1})  with the following  generalized forms with multi-centers: 
\begin{eqnarray}
f&=&\sum_{i=1}^N\frac{M_i}{|{\bm x}-{\bm x_i}|},  \label{eq:fm}\\
g&=&\sum_{i=1}^N\frac{J_iM_i^2(z-z_i)}{2P_iQ_i|{\bm x}-{\bm x_i}|^3}, \label{eq:gm}
\end{eqnarray}
where the constants $M_i,J_i,Q_i\ P_i\ (i=1,\ldots,N)$ denote the mass, angular momentum, electric and magnetic charges of each black holes, with  the following relation at each center of the magnetic and electric charges:
\begin{eqnarray}
\frac{P_i}{2M_i}=\cos^3\alpha,\quad \frac{Q_i}{2M_i}=\sin^3\alpha\ (i=1,2,\ldots N). \label{eq:PiQi}
\end{eqnarray}
Using  Eqs.~(\ref{eq:omega0a}) and (\ref{eq:omega5t}), we can show that the one-forms ${\bm \omega}^0$ and $\tilde {\bm \omega}^5$ take the following explicit forms:
\begin{eqnarray}
{\bm \omega}^0&=&-\sum_{i=1}^N\frac{2J_i}{|{\bm x}-{\bm x_i}|^3}[(y-y_i)dx-(x-x_i)dy],  \label{eq:omega0}\\
 \tilde{\bm \omega}^5&=& \sum_{i=1}^N\frac{2P_i(z-z_i)}{|{\bm x}-{\bm x_i}|}\frac{(y-y_i)dx-(x-x_i)dy}{(x-x_i)^2+(y-y_i)^2}.\label{eq:tomega5}
\end{eqnarray}
Consequently,  the 5D metric for the multi-centered black hole solution takes the form
\begin{eqnarray}
ds^2=\frac{H_-}{H_+}\left[dx^5-\frac{dt}{\sin\alpha}+\frac{1+f \sin^22\alpha}{\sin\alpha\  H_-}(dt+{\bm \omega^0})+\tilde {\bm \omega}^5\right]^2-\frac{1}{H_-}(dt+{\bm \omega^0})^2+H_+d{\bm x}\cdot d{\bm x},
\end{eqnarray}
with Eqs.~(\ref{eq:hpm}), (\ref{eq:fm}), (\ref{eq:gm}), (\ref{eq:omega0}), and (\ref{eq:tomega5}).
\medskip
On the other hand, the 4D metric, gauge field and scalar field after dimensional reduction are expressed as
\begin{eqnarray}
ds^2_{(4)}&=&-\frac{1}{\sqrt{H_+H_-}}(dt+{\bm \omega^0})^2+\sqrt{H_+H_-}d{\bm x}\cdot d{\bm x}, \label{eq:sol:4Dmetric}\\ 
{\bm A}&=&\left(-\frac{1}{\sin\alpha}+\frac{1+f \sin^22\alpha}{\sin\alpha\  H_-}\right)dt+
\frac{1+f \sin^22\alpha}{\sin\alpha\  H_-}{\bm \omega^0}+\tilde {\bm \omega}^5, \label{eq:sol:A}\\
e^{\frac{2\phi}{\sqrt{3}}}&=&\frac{H_+}{H_-}.\label{eq:sol:phi}
\end{eqnarray}
From Eqs.~(\ref{eq:gm}) and (\ref{eq:omega0}), we see that each black hole has either an aligned or anti-aligned spin orientation along the $z$-axis.

\section{Properties of the multi-rotating black hole solution}\label{sec:anaysis}

In this section, we see that this solution is regular and describes asymptotically flat, multi-rotating dyonic black holes, each possessing an extremal horizon.
We demonstrate that curvature singularities are confined inside the horizons and do not occur on or outside them.
Furthermore, we prove the absence of closed timelike curves in the exterior region as well as on the horizons.

\subsection{Near-horizon geometry}

The 4D metric apparently diverges at the points ${\bm x}={\bm x}_i$ but we show that they correspond to smooth Killing horizons, provided the slowly rotating conditions 
$|J_i|<|P_iQ_i|$ for all $i$.
Introducing $r:=|{\bm x}-{\bm x}_i|$, and the standard spherical coordinates $(x,y,z)=(r\sin\theta \cos\phi,r\sin\theta\sin\phi,r\cos\theta)$, we consider the limit of $r\to 0$. 
In this limit, the 4D metric, Maxwell gauge potential and scalar field at $r\to0$ behaves as
\begin{eqnarray}
ds^2_{(4)}&\simeq&-\frac{ r^2}{2\sqrt{P_i^2Q_i^2 -J_i^2 \cos^2\theta }}\left[dt+\frac{2J_i}{r}\sin^2\theta d\phi \right]^2+2\sqrt{P_i^2Q^2_i-J_i^2\cos^2\theta } \left[ \frac{dr^2}{r^2}+ d\Omega^2 \right],\\
{\bm A}&\simeq& -\frac{dt}{\sin\alpha}+2P_i\left(-\cos\theta+\frac{J_i\sin^2\theta }{P_iQ_i-J_i\cos\theta }\right)d\phi,\\
e^{-\frac{2\phi}{\sqrt{3}}} &\simeq&\tan^2\alpha\frac{P_iQ_i-J_i \cos\theta }{P_iQ_i+J_i \cos\theta }, 
\end{eqnarray}
where it should be noted that one can set $A_\theta=0$ and $A_r=0$ at $r=0$ by performing the gauge transformation ${\bm A}\to {\bm A}-d\chi$, with $\chi=\int A_\theta(r=0)\,d\theta + A_r(r=0)\, r$ (here, $A_\theta(r=0)=0$ from the outset).
The metric component $g_{rr}$ diverges  at  $r=0$ but we can show that this is apparent as follows.
In terms of the new coordinates $(v,\phi')$ given by, 
\begin{eqnarray}
dt=dv+\left(\frac{a_0}{r^2}+\frac{a_1}{r} \right)dr,\quad d\phi=d\phi'+\frac{b_0}{r}dr,
\end{eqnarray}
with 
\begin{eqnarray}
a_0&=&\pm2\sqrt{P_i^2Q_i^2-J_i^2},\\
a_1&=&\frac{2M_i^3\sin^42\alpha[1+(\sum_{j\not=i}M_j|{\bm x}_j|^{-1})\sin^22\alpha ]}{a_0},\\
b_0&=&\frac{2J_i}{a_0},
\end{eqnarray}
the apparent divergence can be eliminated. 
It follows that the null surface $r=0$ corresponds to the Killing horizon for the  Killing vector field $\partial/\partial v$. 
The near-horizon limit is defined by $v\to v/\varepsilon$, $r\to\varepsilon r $ and $\varepsilon\to0$, which leads to
\begin{eqnarray}
ds^2_{(4)}&\simeq&-\frac{2f(0)^2\sin^2\theta}{f(\theta)}\left[d\phi'-\frac{rJ_i}{2f(0)^2}dv \right]^2
-\frac{f(\theta)r^2}{2f(0)^2}dv^2
\mp\frac{2f(\theta)}{f(0)}dvdr+2f(\theta)d\theta^2,
\end{eqnarray}
with $f(\theta):=\sqrt{P_i^2Q_i^2-J_i^2\cos^2\theta}$.
This coincides with the near-horizon geometry of the slowly rotating extremal limit of the Rasheed-Larsen black hole with a single horizon.

\subsection{Asymptotic structure}
In terms of the standard spherical coordinates $(x,y,z)=(r\sin\theta \cos\phi,r\sin\theta\sin\phi,r\cos\theta)$, 
the functions, $f,g,H_\pm$ and the $1$-forms ${\bm \omega}^0$, $\tilde {\bm \omega}^5$  behave asymptotically as at $r\to \infty$
\begin{eqnarray}
f&\simeq& \frac{\sum_iM_i}{r}+{\cal O}(r^{-2}),\\
g&\simeq& \sum_i\frac{J_i M_i^2}{2P_iQ_i}\frac{\cos\theta}{r^2}+{\cal O}(r^{-3}),\\
H_\pm&\simeq& 1+\frac{2 (1\pm \cos2\alpha )\sum_i M_i}{r}+{\cal O}(r^{-2}),
\end{eqnarray}
and 
\begin{eqnarray}
{\bm \omega}^0&\simeq& \left(\frac{2\sum_iJ_i}{r}\sin^2\theta+{\cal O}(r^{-2})\right)d\phi,\\
\tilde {\bm \omega}^5&\simeq&\left(-2\sum_i P_i \cos\theta++{\cal O}(r^{-1})\right)d\phi.
\end{eqnarray}

The 5D metric  at $r\to\infty$ behaves as \begin{eqnarray}
ds^2&\simeq&\left(1-\frac{4\cos2\alpha\sum_i M_i }{r}\right) \left[dx^5-2\left(\sum_i P_i\right) \cos\theta d\phi\right]^2+\left(-1+\frac{2\sin^2\alpha \sum_i M_i} {r}\right)\left[dt+\frac{2\sum_iJ_i}{r}\sin^2\theta d\phi \right]^2\notag\\
&&+\left(1+\frac{2\cos^2\alpha \sum_i M_i} {r}\right)\left[dr^2+r^2\left(d\theta^2+\sin^2\theta d\phi^2\right)\right], 
\end{eqnarray}
indicating that the spacetime asymptotically approaches an $S^1$ fiber bundle over 4D Minkowski space.
On the other hand, at $r \to \infty$, the dimensionally reduced 4D fields behave as follows:
\begin{eqnarray}
ds^2_{(4)}&\simeq&\left(-1+\frac{2 \sum_i M_i} {r}\right)\left[dt+\frac{2\sum_iJ_i}{r}\sin^2\theta d\phi \right]^2+\left(1+\frac{2 \sum_i M_i} {r}\right) \left[ dr^2+r^2 (d\theta^2+\sin^2 \theta d\phi^2) \right]\\
{\bm A}&\simeq& \frac{-2\sum_iQ_i }{ r}dt-2\left(\sum_iP_i \right)\cos\theta d\phi,\\
\phi &\simeq& \frac{2\sqrt{3} \cos2\alpha \sum_i M_i}{r}.
\end{eqnarray}
Thus, the dimensionally reduced spacetime is asymptotically flat.
The ADM mass, ADM angular momentum, and the total electric and magnetic charges are given respectively by
\begin{eqnarray}
&&M=\sum_{i=1}^NM_i,\quad J=\sum_{i=1}^NJ_i,
\end{eqnarray}
\begin{eqnarray}
&&P=\sum_{i=1}^NP_i=2M\cos^3\alpha,\quad Q=\sum_{i=1}^N Q_i=2M \sin^3\alpha,
\end{eqnarray}
where each individual electric and magnetic charge satisfies
\begin{eqnarray}
(P_i,Q_i)=2M_i (\cos^3\alpha,\sin^3\alpha),
\end{eqnarray}
leading to the identity
\begin{eqnarray}
\left(\frac{P}{2M_i}\right)^{\frac{2}{3}}+\left(\frac{Q}{2M_i}\right)^{\frac{2}{3}}=1.\label{eq:ratio}
\end{eqnarray}

\subsection{Regularity}
If curvature singularities exist on or outside the horizons, they appear at points where the metric or its inverse diverges, which happens only on the surfaces $H_+(x,y,z)=0$ or $H_-(x,y,z)=0$.
Indeed, since the Kretschmann scalar can be calculated  as
\begin{eqnarray}
R^{(4)}_{\mu\nu\rho\lambda}R^{(4)\mu\nu\rho\lambda}\sim \frac{1}{(H_+H_-)^3},
\end{eqnarray}
curvature singularities arise precisely on  the surfaces where either  $H_+(x,y,z)=0$ or $H_-(x,y,z)=0$. 
We can show that such singularities do not exist on and outside the event horizons at ${\bm x}={\bm x}_i$ provided that $|J_i|<|P_iQ_i|$.
To demonstrate this, it is sufficient to verify that $H_\pm>0$ on and outside the horizons, since asymptotically $H_\pm \to 1>0$ as $r\to\infty$.
First, we note that under the assumptions $M_i>0$, we have $f>0$, while $g$ can take both signs. 
Using this, we write for $(c_2,s_2):=(\cos2\alpha,\sin2\alpha)$, the functions $H_\pm$ as
\begin{eqnarray}
H_\pm   &=& 1+2(1\pm c_2)f+s_2^2(1\pm c_2) f^2+2s_2^2(c_2\pm1)g\notag\\
              &> & s_2^2(1\pm c_2)+2s_2^2(1\pm c_2)f+s_2^2(1\pm c_2) f^2+2s_2^2(c_2\pm1)g\mp s_2^2c_2\notag\\
              &= &s_2^2(1\pm c_2 )[(1+f)^2\pm2g]\mp s_2^2c_2\notag\\
              &\ge &s_2^2(1\pm c_2 )[(1+f)^2- 2|g|]\mp s_2^2c_2\notag\\
              &>&s_2^2(1\pm c_2 )\mp s_2^2c_2\notag\\
              &=&s_2^2\notag\\
              &>&0,
\end{eqnarray}
where we have used the inequality
\begin{eqnarray}
(1+f)^2-2|g|&=&\left(1+\sum_if_i\right)^2-2|\sum_i g_i|\notag\\
                  &>&1+\sum_if_i^2-2\sum_i |g_i|\notag\\
                  &>&1+\sum_i M_i^2\left[\frac{1-|J_i/P_iQ_i|}{|{\bm x}-{\bm x}_i|^2}\right]\notag\\
                  &>1&,
\end{eqnarray}
with
\begin{eqnarray}
f_i:=\frac{M_i}{|{\bm x}-{\bm x_i}|},\quad g_i:=\frac{J_iM_i^2(z-z_i)}{2P_iQ_i|{\bm x}-{\bm x_i}|^3}.
\end{eqnarray}

\subsection{Absence of CTCs}

Here, we show the nonexistence of CTCs everywhere on and outside the horizons, provided that the inequality $|J_i| < |P_i Q_i|$ holds for each $i = 1, \ldots, N$.
The condition for the absence of CTCs is equivalent to requiring that the $2$D metric $g_{(2)IJ}dx^I dx^J$ $(x^I, x^J = x, y)$ is positive definite everywhere on and outside the horizons. This condition can be expressed in terms of its trace and determinant as follows:
\begin{eqnarray}
{\rm tr\ }g_{(2)} &=& g_{xx} + g_{yy} = \frac{2H_+ H_- - (\omega^0_x)^2 - (\omega^0_y)^2}{\sqrt{H_+ H_-}} > 0, \label{eq:tr} \\
{\rm det\ }g_{(2)} &=& g_{xx} g_{yy} - g_{xy}^2 = H_+ H_- - (\omega^0_x)^2 - (\omega^0_y)^2 > 0.\label{eq:det}
\end{eqnarray}
Since $H_+ H_- > 0$, it is straightforward to verify that if the determinant condition~\eqref{eq:det} is satisfied, then the trace condition~\eqref{eq:tr} is also automatically satisfied. Therefore, it is sufficient to consider only the inequality~\eqref{eq:det}.

\medskip
The determinant can be expressed as
\begin{eqnarray}
{\rm det\ }g_{(2)}&=&1+4f+s_2^2(6f^2+4c_2g)+4s_2^4f^3+s_2^6(f^4-4g^2)-(\omega^0_x)^2 - (\omega^0_y)^2\\
&>&1+\sum_i[s_2^6(f_i^4-4g_i^2)-(\omega^0_{xi})^2-(\omega^0_{yi})^2]\notag\\
&&+2\sum_{i\not= j}[s_2^6(3f_i^2f_j^2-4g_ig_j)-(\omega^0_{xi})(\omega^0_{xi})-2(\omega_{iy})(\omega_{jy})],
\end{eqnarray}
with
\begin{eqnarray}
\omega^0_{xi}=\frac{2J_i(y-y_i)}{|{\bm x}-{\bm x_i}|^3},\quad 
\omega^0_{yi}=\frac{-2J_i(x-x_i)}{|{\bm x}-{\bm x_i}|^3},
\end{eqnarray}
where the second and third summations can be shown to be positive under the conditions $|J_i|<|P_iQ_i|\ (i=1,\ldots,N)$. Indeed,
\begin{eqnarray}
s_2^6(f_i^4-4g_i^2)-(\omega^0_{xi})^2-(\omega^0_{yi})^2&=&s_2^6\left(\frac{M_i^4}{|{\bm x}-{\bm x_i}|^4}-\frac{J_i^2M_i^4(z-z_i)^2}{P_i^2Q_i^2|{\bm x}-{\bm x_i}|^6}\right)
-\frac{4J_i^2(y-y_i)^2}{|{\bm x}-{\bm x_i}|^6}
-\frac{4J_i^2(x-x_i)^2}{|{\bm x}-{\bm x_i}|^6}\notag\\
&=&s_2^6\left(\frac{M_i^4}{|{\bm x}-{\bm x_i}|^4}-\frac{J_i^2M_i^4(z-z_i)^2}{P_i^2Q_i^2|{\bm x}-{\bm x_i}|^6}\right)
-s_2^6\frac{J_i^2M_i^4[(x-x_i)^2+(y-y_i)^2]}{P_i^2Q_i^2|{\bm x}-{\bm x_i}|^6}\notag\\
&=&s_2^6\frac{M_i^4(1-|J_i/P_iQ_i|^2)}{|{\bm x}-{\bm x_i}|^4}\notag\\
&>&0,
\end{eqnarray}
and
\begin{eqnarray}
&&s_2^6(3f_i^2f_j^2-4g_ig_j)-(\omega^0_{xi})(\omega^0_{xj})-(\omega^0_{yi})(\omega^0_{yj})\notag\\
&&=s_2^6\left(\frac{3M_i^2M_j^2}{|{\bm x}-{\bm x_i}|^2|{\bm x}-{\bm x_j}|^2}-\frac{J_iJ_j M_i^2M_j^2(z-z_i)(z-z_j)}{P_iP_jQ_iQ_j|{\bm x}-{\bm x_i}|^3|{\bm x}-{\bm x_i}|^3}\right)-\frac{4J_iJ_j[(x-x_i)(x-x_j)+(y-y_i)(y-y_j)]}{|{\bm x}-{\bm x_i}|^3|{\bm x}-{\bm x_j}|^3}\notag\\
&&=s_2^6\left(\frac{3M_i^2M_j^2}{|{\bm x}-{\bm x_i}|^2|{\bm x}-{\bm x_j}|^2}-\frac{J_iJ_j M_i^2M_j^2(z-z_i)(z-z_j)}{P_iP_jQ_iQ_j|{\bm x}-{\bm x_i}|^3|{\bm x}-{\bm x_i}|^3}\right) \notag\\
&&\hspace{5cm}-s_2^6\frac{J_iJ_jM_i^2M_j^2[(x-x_i)(x-x_j)+(y-y_i)(y-y_j)]}{P_iP_jQ_iQ_j{|\bm x}-{\bm x_i}|^3|{\bm x}-{\bm x_j}|^3}\notag\\
&&=s_2^6\left(\frac{3M_i^2M_j^2}{|{\bm x}-{\bm x_i}|^2|{\bm x}-{\bm x_j}|^2}-\frac{J_iJ_j M_i^2M_j^2({\bm x-{\bm x_i}})\cdot({\bm x}-{\bm x_j})}{P_iP_jQ_iQ_j|{\bm x}-{\bm x_i}|^3|{\bm x}-{\bm x_i}|^3}\right)\notag\\
&&>s_2^6\left(\frac{3M_i^2M_j^2}{|{\bm x}-{\bm x_i}|^2|{\bm x}-{\bm x_j}|^2}-\frac{|J_i||J_j| M_i^2M_j^2}{|P_i||P_j||Q_i||Q_j||{\bm x}-{\bm x_i}|^2|{\bm x}-{\bm x_i}|^2}\right)\notag\\
&&>s_2^6\frac{3M_i^2M_j^2(1-|J_i/P_iQ_i||J_j/P_jQ_j|)}{|{\bm x}-{\bm x_i}|^2|{\bm x}-{\bm x_j}|^2}\notag\\
&&>0.
\end{eqnarray}
The positivity of these equations implies that ${\rm det\ } g_{(2)} > 1$.
Therefore, no CTCs exist on or outside the horizons.

\subsection{Parameter region for physical solutions}\label{sec:limit}

From Eq.~(\ref{eq:PiQi}) and the slowly rotating condition $|J_i|<|P_iQ_i|$, it follows that when $\alpha=0$ or $\alpha=\pi/2$—corresponding to $Q_i=0$ or $P_i=0$, respectively—the angular momenta $J_i$ must vanish.
In these cases, the horizon areas shrink to zero and the solutions become singular.
Therefore, we restrict our analysis to the range $0<\alpha<\pi/2$.

 \medskip
 As discussed in the subsections, 
 the regularity and the absence of CTCs on and outside the horizons require that the parameters must be subjected to
\begin{eqnarray}
M_i>0,\quad |J_i|<P_iQ_i,\quad 0<\alpha<\frac{\pi}{2}.
\end{eqnarray}
 When $\alpha\to \pi/4$, the electric and magnetic charges become equal, $Q_i=P_i= \sqrt{2}M_i$ 
for all $i$, and then the solution exactly coincides with the Teo-Wan solution~\cite{Teo:2023wfd}.
In addition, when $J_i=0$ for all $i$, the scalar field vanishes, and the solution reduces to the static Majumdar-Papapetrou black hole solution with equal electric and magnetic charges in Einstein-Maxwell theory~\cite{Matsuno:2012hf}. 
On the other hand, when $J_i=0$ for all $i$ but $\alpha\neq\pi/4$, the scalar field does not vanish, and the solution reduces to a static dyonic multi-black hole configuration with a scalar field.

\section{Summary and Discussion}\label{sec:summary}

We have constructed multi-Kaluza-Klein black hole solutions in 5D Einstein gravity that, upon dimensional reduction, yield multi-centered rotating black holes with both electric and magnetic charges in 4D Einstein-Maxwell-dilaton theory.
In this work, we have extended the multi-centered rotating black hole solution of Teo and Wan to the case of unequal electric and magnetic charges.
 The resulting spacetime has remained regular---free from curvature singularities, conical defects, Dirac-Misner strings, and closed timelike curves---both on or outside the horizons, provided the black holes have had either aligned or anti-aligned spin orientations. 
Furthermore, the solutions has important limits to physical solutions: 
when $J_i=0$ for all $i$, the solution reduces to the static multi-black holes with a scalar field, 
in addition, when $Q_i=P_i=M_i/\sqrt{2}$ for all $i$, the scalar field vanishes, and the solution reduces to the static Majumdar-Papapetrou black hole solution with equal electric and magnetic charges in Einstein-Maxwell theory~\cite{Matsuno:2012hf}.

\medskip
We comment on the physical features of the present solution as follows:
(1) Each black hole carries nonzero angular momentum, whereas each horizon angular velocity vanishes. This should not  be surprising, since the rotations of the Maxwell field and the scalar field around the horizons contribute to the angular momenta.
(2) Interpreted as a 5D solution with a compact extra dimension, the geometry also describes 5D multi-rotating black holes. The spatial cross-section of each horizon can be regarded as a Hopf bundle, an $S^1$ fiber bundle over an $S^2$ base. The electric and magnetic charges correspond, respectively, to the momentum along the fifth dimension and to the twist between that direction and the 4D spacetime.

\medskip
In this paper, we have focused on 5D Einstein theory with a compact dimension.
However, we expect that this method can be applied more broadly to any theory whose coset matrix is symmetric, such as supergravity.
As a future direction, we aim to derive extremal black hole solutions in a variety of 4D theories which, after Kaluza-Klein reduction, can be reformulated in terms of 3D  gravity coupled to a sigma model with a symmetric target space~\cite{Gaiotto:2007ag}.
For instance, in the bosonic sector of 5D minimal supergravity with two commuting Killing vectors (one timelike and one spacelike), the relevant coset space is $G_{2(2)}/[SL(2,{\mathbb R})\times SL(2,{\mathbb R})]$.
Since the most general black hole solution with six independent charges has been constructed in Ref.~\cite{Tomizawa:2012nk}, it may be possible to further extend this framework to construct multi-centered rotating black hole solutions.

\acknowledgments
ST was supported by JSPS KAKENHI Grant Number 21K03560.
RS was supported by JSPS KAKENHI Grant Number JP18K13541.




\end{document}